\documentclass[preprint,showpacs,nofootinbib,prc,aps]{revtex4}

\usepackage{graphicx,amstext,amssymb}

\begin{document}

\title{Matching of nonthermal initial conditions and hydrodynamic
stage in ultrarelativistic heavy-ion collisions}

\author{S.V. Akkelin}
\author{ Yu.M. Sinyukov}
\affiliation{Bogolyubov Institute for Theoretical Physics,
Metrolohichna str. 14b, 03680 Kiev,  Ukraine}

\begin{abstract}

A simple approach is proposed allowing actual calculations of the preequilibrium dynamics in ultrarelativistic
heavy-ion collisions to be performed for a far-from-equilibrium initial state. The method is based on the
phenomenological macroscopic equations that describe the relaxation dynamics of the energy-momentum tensor
and are motivated by Boltzmann kinetics in the relaxation-time approximation. It gives the possibility to match
smoothly a nonthermal initial state to the hydrodynamics of the quark gluon plasma. The model contains two
parameters, the duration of the prehydrodynamic stage and the initial value of the relaxation-time parameter, and
allows one to assess the energy-momentum tensor at a supposed time of initialization of the hydrodynamics.

\end{abstract}


 \maketitle

 \section{Introduction}

A comprehensive analysis of the experimental data from
the Relativistic Heavy-Ion Collider (RHIC) has shown that a
quark-gluon plasma (QGP) is created in these collisions, and
that the thermalized QGP is, perhaps, the most perfect liquid
possible in nature. This conclusion is based, in particular,
on the success of ideal hydrodynamics in describing of the
basic features of heavy-ion collisions at RHIC energies (for a review see, e.g.,
Ref. \cite{Heinz-1}). Recently, essential progress has
been made in the development and applications of viscous
hydrodynamics for RHIC heavy-ion collisions (for reviews and
recent results see, e.g., \cite{Heinz-1,Pratt-1,Teaney} and
references therein). Viscous
hydrodynamics accounts for deviations from local equilibrium
by means of dissipative transport coefficients, and, therefore,
has a more extended region of applicability as compared to
ideal hydrodynamics. Nevertheless, it should be emphasized
that the domain of validity of viscous hydrodynamics is still the
hydrodynamic regime; thus hydrodynamics is valid when the
relaxation time $\tau_{\text{rel}}$ is much
smaller than the inverse expansion rate, $1/\partial_{\mu}
u^{\mu}$, that is,  $\tau_{\text{rel}}\partial_{\mu} u^{\mu} \ll 1$,
and one can expect that  hydrodynamics breaks down when
$\tau_{\text{rel}}\partial_{\mu} u^{\mu} \sim 1/2$. In practice,
hydrodynamics breaks down in ultrarelativistic heavy-ion
collisions at the very initial nonequilibrium stage, near the edge
of the fireball, and at the later rarefied kinetic stage of matter
evolution. Since hydrodynamics is based on the assumption
that the system is near local thermal equilibrium
\cite{Huang,Balescu,Zubarev}, it
is natural that its region of applicability cannot be reliably
determined from within itself, and can be properly estimated
only with the help of an appropriate nonequilibrium theory.

In addition to the breakdown times, one also needs to
specify initial conditions, such as the energy density, fluid
velocities, and viscous shear tensor to apply hydrodynamics.
Evidently, the initial conditions for hydrodynamics are determined
at the very initial nonequilibrium stage of the matter
evolution in ultrarelativistic heavy-ion collisions. This stage
is well understood now based on the color glass condensate
(CGC) approach (for a review see, e.g., Ref.
\cite{CGC}), which
is adequate at RHIC and, probably, Large Hadron Collider
(LHC) energies, although an explanation of the thermalization
and nearly perfect fluidity provides a challenge for the theory of quark-gluon
matter (for  recent results, see Refs. \cite{Kovchegov,Zxu-1}). It
is worth noting, however, that the assumption of very early
thermalization (say, $\tau_{\text{th}} = 0.3$ fm/c) is not necessary for data
description in hydrodynamics, since the transverse collective
flows and their azimuthal anisotropy for noncentral collisions
can appear already at the beginning of the hydrodynamic
expansion as a result of the development of the transverse
velocities at the prethermal (glasma \cite{Glasma} or partonic or string
\cite{Werner}) stage \cite{flow}. That is why an analysis of matter
evolution at the prethermal stage, which determines the further
hydrodynamic expansion, is so important..

 The nonthermal initial conditions are related typically to the Bjorken  proper time $\tau_0$
when  the system can be characterized by the phase-space density of
individual partons. In the CGC approach this "formation time" was
estimated to be
 $\tau_0 \approx 0.1-0.3$ fm/c for central
collisions in midrapidity at  LHC  and  RHIC energies
\cite{CYM-1,Lappi}. Taking into account the theoretical estimate
of the thermalization time scale $\tau_{\text{th}}=1-1.5$ fm/c for
the LHC and RHIC  heavy-ion collisions
 \cite{Zxu-1}, one can hardly
expect hydrodynamics to be applicable at  $\tau =\tau_{0} < 1$ fm/c, because then
$\tau_{\text{rel}}\partial_{\mu} u^{\mu} \sim \tau_{\text{th}}/\tau
> 1 $.

Thus, to obtain the initial conditions for hydrodynamics at $\tau =
\tau_{\text{th} }\gtrsim 1$ fm/c, one needs to match the very early
initial stage of the nuclei collisions, when the hydrodynamic
approximation is outside the regime of its validity, with an
almost local equilibrium state of the QGP. It is noteworthy that
phase-space distributions derived from the CGC approach are
highly anisotropic, and so the distribution functions are quite
nonequilibrium. The longitudinal momentum distribution is
much more narrow than transverse one, and asymptotically, at
late times, it is a delta function $\delta (p_{z})$ at $z=0$  \cite{Larri}. The latter
approximation is often used for a description of the initial stage
in \textit{A }+ \textit{A} collisions (see, e.g.,
\cite{Zxu-2}).\footnote{ One should understand, however, that such
a utilization of the asymptotic approximation  is not quite correct
from the point of view of quantum theory. Indeed, the Wigner function
$f_{\text{W}}(x,p)$, which  is the quantum mechanical analog of the
classical phase-space density $f(x,p)$, satisfies the restriction $
\int f_{\text{W}}^{2}(x,p)d^{3}pd^{3}x \leq (2 \pi \hbar)^{-3}$ (see,
e.g., \cite{QM}): here the normalization condition $\int
f_{\text{W}}(x,p)d^{3}pd^{3}x =1 $ is supposed. Then, in order to
escape a contradiction with quantum mechanics, smooth
boost-invariant prescriptions for the longitudinal part of the 
distribution $f(x,p)$ from Refs. \cite{Sin-1} or \cite{Flor-1}
should be used for a description of the initial stage in \textit{A
}+ \textit{A} collisions.}

 In this paper, we formulate phenomenological macroscopic
equations accounting for the energy-momentum and quantum
number conservation laws which allow one to connect the
arbitrary initial, probably highly nonequilibrium state of the
system with the (partially) equilibrated thermal state, when
one can start to use (viscous) hydrodynamics to describe the
further matter evolution.\footnote{A similar effort was
undertaken in Ref. \cite{Strickland}, where the preequilibrium dynamics
of a $(0+1)$-dimensional QGP was matched to second-order viscous
hydrodynamics.} The model exploits a minimal set of phenomenological
parameters and can be utilized to assess the initial conditions for
hydrodynamic expansion, which can be used,  then,  for a hydrodynamic
modeling of ultrarelativistic heavy-ion collisions.

\section{Energy and momentum relaxation dynamics for the early stage
of  ultrarelativistic heavy-ion collisions}

It is well known that statistical derivation of the macroscopic
hydrodynamic equations for gases is based on an approximate
solution of the Boltzmann equations near the local equilibrium
distribution (see, e.g., Refs. \cite{Huang,Balescu,Zubarev}). The form
of hydrodynamic equations derived can then be spread out
phenomenologically from a gas to dense fluids if they are
near local thermal equilibrium.More precisely, hydrodynamic
equations for dense systems like liquids near local thermal
equilibrium can be derived by means of the (quasi) equilibrium
statistical operator method based on the conditions of a
maximum of local entropy and under appropriate auxiliary
conditions \cite{Zubarev}. The proximity to
local thermal equilibrium allows one to utilize for a description of
these  inhomogeneous evolutional systems the Gibbs thermodynamic
relations in the  same  form as for equilibrium systems
\cite{Balescu}, and to get the equations of viscous hydrodynamics in
the closed form. However, if the system is far from local
equilibrium, it can be described macroscopically only through the
unclosed hierarchical structure of the hydrodynamic balance
equations expressing the conservation laws in the system
\cite{Balescu}. All the momenta of the physical values are present
there, and they cannot be calculated without knowledge of the
evolution of the nonequilibrium distribution function \cite{Huang}.
Although the second-order viscous hydrodynamic equations increase
 precision compared with the first-order equations and, moreover,
have attractive features from the point of view of mathematical
formulation,  the domain of applicability of these equations is
still the hydrodynamic domain: the microscopic scales like the mean
free path are much smaller than the macroscopic dimensions (the system's
homogeneity lengths). Because of  this,  one can conjecture that
dissipative fluid dynamics is inapplicable in its standard form to the 
very initial stage, $\tau < 1 $ fm, of ultrarelativistic heavy-ion
collisions (see also Ref. \cite{Baier}), when the rate of expansion
is too high and the interactions are not strong enough to validate the 
application of the hydrodynamic approach. Thereby, the very initial
nonthermal state (glasma?) of matter evolution in \textit{A} +
\textit{A} collisions cannot be described hydrodynamically, unlike
the subsequent thermal QGP stage, where hydrodynamics seems to work
well \cite{Heinz-1}.\footnote{Note that  the results of viscous
hydrodynamic calculations depend on exact values of  the transport
coefficients that still cannot be derived
 unambiguously  from  the  complicated  theory of  microscopic dynamics.
At the later, low-density gaslike stage,  the viscous hydrodynamics
can be coupled to  the Boltzmann kinetics of the hadron resonance gas
\cite{Pratt-2}.}

To perform the heuristic arguments to derive
phenomenological macroscopic equations for a description of
transient prehydrodynamic behavior, let us first assume that the
initial phase-space density is associated with the partons,while
the resulting expressions can be used as a phenomenological
approach not only for partonic matter but also for other forms
of prethermal matter like glasma and strings. Usually, if the
exact dynamics is cumbersome or unknown, establishment
of the local equilibrium state in the system is modeled by
the kinetic equation for the phase-space distribution function $f(x,p)$ in the relaxation time
approximation, with $f^{\text{l eq}}(x,p)$ as the target function to which the phase-space distribution
tends. The corresponding equation has the form
\begin{eqnarray}
 \frac{p^{\mu} \partial f(x,p)}{\partial x^{\mu }}
 =-p^{\mu}u_{\mu} \frac{f(x,p)-f^{\text{l
 eq}}(x,p)}{\tau^*_{\text{rel}}(x,p)}, \label{boltz-1}
\end{eqnarray}
where $\tau^*_{\text{rel}}$ is the relaxation-time parameter in the
local rest frame of the energy flow  (in general it is some function
of $(x,p)$), and $u^{\mu}(x)$ is the four-vector energy flow field,
\begin{eqnarray} u^{\mu}=\frac{T^{\mu
\nu}u_{\nu}}{T^{\mu\nu}u_{\mu}u_{\nu}}=\frac{T^{\mu
\nu}u_{\nu}}{\epsilon}, \label{011} \\
T^{\mu \nu}(x)=\int d^{3}p \frac{p^{\mu}p^{\nu}}{p_{0}}f(x,p).
\label{2}
\end{eqnarray}
To determine  the parameters (temperature, etc.) that  define the local
equilibrium state, $f^{\text{l eq}}(x,p)$, it is necessary to
implement the energy-momentum conservation equations,
\begin{eqnarray}
\partial_{\mu}T^{\mu \nu}(x)=0, \label{tensor-1}
\end{eqnarray}
and, if necessary, the conservation equations for (net) quantum
numbers $q_i$,
\begin{eqnarray}
\partial_{\mu}q_{i}^{\mu}(x)=0,
\label{cons-2}
\end{eqnarray}
with current $q_{i}^{\mu}(x)$,
\begin{eqnarray}
q_{i}^{\mu}(x)=\int
\frac{d^{3}k}{k_{0}}k^{\mu}[f_{q_{i}}(x,k)-f_{-q_{i}}(x,k)].
\label{number-def1}
\end{eqnarray}
Then Eqs. (\ref{tensor-1}) and  (\ref{cons-2})
 together with the formal  solution  of Eq.
(\ref{boltz-1}),
\begin{eqnarray}
f(t,{\bf r},p) =f(t_{0},{\bf r}-\frac{{\bf p}}{p_{0}}
(t-t_{0}),p)P(t_{0},t,\textbf{r},p) + \nonumber \\
\stackrel{t}{%
%TCIMACRO{\underset{t_{0}}{\int }}%
%BeginExpansion
\mathrel{\mathop{\int }\limits_{t_{0}}}%
%EndExpansion
}f^{\text{l eq}}(t^{\prime },{\bf r}-\frac{{\bf
p}}{p_{0}}(t-t^{\prime }),p) \frac{d}{dt^{\prime
}}P(t^{\prime},t,\textbf{r},p) dt^{\prime},
 \label{boltz-3-1}
\end{eqnarray}
where the probability for particle with momentum $\textbf{p}$ to
propagate freely from point $(t^{\prime},{\bf r})$ to point $(t,{\bf
r}+\frac{{\bf p}}{p^0}(t-t^{\prime}))$ is
\begin{equation}
P(t^{\prime},t,\textbf{r},p)=
\exp \left\{ -\stackrel{t}{%
%TCIMACRO{\underset{t^{\prime }}{\int }}%
%BeginExpansion
\mathrel{\mathop{\int }\limits_{t^{\prime }}}%
%EndExpansion
}\frac{p_{\mu}u^{\mu}(s,{\bf r}-\frac{{\bf
p}}{p_{0}}(t-s))}{p_{0}\tau^*_{\text{rel}}(s,{\bf r}-\frac{{\bf
p}}{p_{0}}(t-s),p)}ds\right\},  \label{boltz-3-2}
\end{equation}
lead to very complicated equations for the thermal and hydrodynamic
fields (temperature, collective velocities, etc.) that define the
target local equilibrium state described by $f^{\text{l eq}}(x,p)$.
The problem becomes even more severe if one takes into account that
$\tau^*_{\text{rel}}$ is a function of $f^{\text{l eq}}$. Also
note that the target state is reached in a finite time interval at
$t = t_{\text{th}},$\footnote{Of course, the same statement is easy to
reformulate for the case when one operates with an arbitrary
three-dimensional spacelike "thermalization" hypersurface  where
$t=t_{\text{th}}(\textbf{r})$ instead of a constant time
$t_{\text{th}}$.} only if the relaxation-time parameter in Eq.
(\ref{boltz-1}) vanishes at $t\rightarrow t_{\text{th}}$:
$\tau^*_{\text{rel}}(t\rightarrow t_{\text{th}},\textbf{r},p)
\rightarrow 0$.

All this makes difficult a utilization of Eq. (\ref{boltz-1}) for a
matching of a far-from-equilibrium  initial state to perfect or
viscous hydrodynamics with presupposed transport
coefficients\footnote{Then the target function in Eq.
(\ref{boltz-1}) could have a nonequilibrium form with parameters
that are linked to transport coefficients.} in relativistic
heavy-ion collisions. The situation reflects the fact that the
solutions of the Boltzmann equations as a rule do not correspond to
any hydrodynamics, or, in some conditions (in the vicinity of a local
equilibrium state), correspond to solutions of viscous hydrodynamic
equations with a very specific form of the viscosity coefficients
defined by the cross sections or by the relaxation-time parameters
in the relaxation-time approximation of the collision terms. To
utilize this in practice a quite complicated Chapman-Enskog method
(see, e.g., \cite{Huang,Groot}) is applied. This method is
inapplicable, however, if the system is initially nonthermal.

 Here we propose a simple phenomenological method
allowing one to describe the relaxation of an initially
far-from-equilibrium system toward the neighborhood of a
local equilibrium state. The latter is described by the target
energy-momentum tensor and is associated, in general, with
viscous hydrodynamics. For simplicity, we will consider
in detail a model with the target energy-momentum tensor
associated with a perfect fluid; a generalization for the viscous
target energy-momentum tensor is straightforward. To arrive at
such a model, it is useful to start again with the relaxation-time
approximation (\ref{boltz-1}) which 
has the formal solution (\ref{boltz-3-1}), (\ref{boltz-3-2}). Then,
 assuming that the maximum of the integrand in
(\ref{boltz-3-1})   occurs  at the  upper limit of the integral
and, moreover, that $f^{\text{l eq}}(x,p)$ is a relatively smooth
function, one can   factor out from the integral in
(\ref{boltz-3-1}) the function $f^{\text{l eq}}(t^{\prime },{\bf
r}-\frac{{\bf p}}{p_{0}}(t-t^{\prime }),p)$ at point $t^{\prime
}=t$ and arrive at the  simple form,
\begin{eqnarray}
f(x,p) =f(t_{0},{\bf r}-\frac{{\bf p}}{p_{0}} (t-t_{0}),p)
P(t_{0},t,\textbf{r},p) + f^{\text{l eq}}(t,{\bf
r},p)(1-P(t_{0},t,\textbf{r},p)),
 \label{boltz-3-55}
\end{eqnarray}
which  demonstrates an explicit transition to a local equilibrium state
at $t=t_{\text{th}}$ if $P(t_{0},t_{\text{th}},\textbf{r},p)=0$ and
is, of course, just an approximation of (\ref{boltz-3-1}). Let us
write this expression  in some more general relativistic invariant
form,  aiming to use the curvature coordinates in our further
analysis,
\begin{eqnarray}
f(x,p) = f_{\text{free}}(x,p){\cal P}(x,p) + f^{\text{l
eq}}(x,p)(1-{\cal P}(x,p)),
 \label{boltz-3-5}
\end{eqnarray}
where $f_{\text{free}}(x,p)$ is the distribution function of freely
streaming partons,
\begin{eqnarray}
p^{\mu}\partial_{\mu}f_{\text{free}}(x,p) = 0,
 \label{free-1}
\end{eqnarray}
which  coincides with $f(x,p)$  at some  three-dimensional
spacelike hypersurface where initial conditions for $f(x,p)$ are
formulated, for example, $f_{\text{free}}(x,p)=f(t_{0},{\bf r}-\frac{{\bf
p}}{p_{0}} (t-t_{0}),p)$  for the initial distribution function
$f(t_{0},{\bf r},p)$. Here ${\cal P}(x,p)$  has the same probability
interpretation as $P$: it is the probability for the particle
emitted on the initial hypersurface (in simplest case - at time $t_{0}$)    with momentum $\textbf{p}$ 
to reach point $x$:
\begin{eqnarray}
{\cal P}(x,p)= \exp \left\{ - \int d^{4}x' G_{p}(x-x')
\frac{p^{\mu}u_{\mu}(x')}{\tau^*_{\text{rel}}(x',p)} \right\},
\label{p-def} \\
p^{\mu}\partial_{\mu}G_{p}(x-x')=\delta^{(4)}(x-x'), \label{green1}
\end{eqnarray}
  ${\cal P}=1$  initially at $t=t_0$ (or at some
three-dimensional hypersurface $t_{0}(\textbf{r})$ where initial
conditions are specified),  ${\cal P}= 0$  at $t = t_{\text{th}}$
(or at some hypersurface $t_{\text{th}}(\textbf{r})$) when
$\tau^*_{\text{rel}}\rightarrow 0$. One can easily  see that
(\ref{boltz-3-5}) satisfies the equation
 \begin{eqnarray}
 \frac{p^{\mu} \partial f(x,p)}{\partial x^{\mu }}
 =-p^{\mu}u_{\mu} \frac{f(x,p)-f^{\text{l
 eq}}(x,p)}{\tau^*_{\text{rel}}(x,p)}+ \frac{p^{\mu} \partial f^{\text{l
 eq}}(x,p)}{\partial x^{\mu }}(1-{\cal P}(x,p)).
 \label{boltz-2}
\end{eqnarray}
Note here that, while Eq. (\ref{boltz-3-5})  is relatively simple,
the equations for $f^{\text{l eq}}(x,p)$ that  follow from
conservation laws (\ref{tensor-1}) and  (\ref{cons-2}) are, in general,
very  complicated (even if one does not relate $\tau^*_{\text{rel}}$
with $f^{\text{l eq}}(x,p)$) because of the momentum dependence of
${\cal P}(x,p)$.

Our idea is to utilize the approximate formal solution
(\ref{boltz-3-5}), which  preserves the most important properties of
the true dynamics at the prethermal stage of evolution, to
 describe the relaxation dynamics of the energy-momentum tensor. To do it,
we first make the  approximations
\begin{eqnarray}
\tau^*_{\text{rel}}(x,p)\approx \tau^*_{\text{rel}}(x), \quad
 {\cal P}(x,p)\approx {\cal P}(x), \label{appr-2}
\end{eqnarray}
which  significantly simplifies modeling of the relaxation dynamics at
the early prethermal stage of evolution in ultrarelativistic
heavy-ion collisions,\footnote{The example of explicit representation
of ${\cal P}(x)$ will be given
in the next section in hyperbolical Bjorken coordinates for constant
longitudinal proper times corresponding to initial and final ("thermalization")
hypersurfaces.} because then, in particular, the energy-momentum
tensor (\ref{2}) takes the simple form
\begin{eqnarray}
T^{\mu \nu}(x)=T^{\mu \nu}_{\text{free}}(x){\cal
P}(x)+T_{\text{hyd}}^{\mu \nu}(x)(1-{\cal P}(x)),
\label{green2}\\
T^{\mu \nu}_{\text{free}}(x)=\int d^{3}p \frac{p^{\mu}p^{\nu}}{p_{0}}f_{\text{free}}(x,p), \label{green3} \\
T^{\mu \nu}_{\text{hyd}}(x)=\int d^{3}p
\frac{p^{\mu}p^{\nu}}{p_{0}}f^{\text{l eq}}(x,p). \label{green4}
\end{eqnarray}
Even though  (\ref{green2}) is a rather rough approximation for the true
energy-momentum tensor, it preserves the desired properties of
the true expression, namely, it demonstrates a transition from
a far-from-equilibrium state to a local equilibrium state and,
which is very important, it allows one to account for the energymomentum
conservation laws and, if necessary, quantum
number conservation laws in a simple form. That is, accounting
for the energy-momentum conservation laws $\partial_{\mu}T^{\mu \nu}(x)=0$,
with $T^{\mu \nu}(x)$ defined in (\ref{green2}), and taking into
account that $\partial_{\mu}T^{\mu \nu}_{\text{free}}(x)=0$, we
arrive to the following equations:\footnote{Note here, to avoid
misunderstanding, that while the expression (\ref{green2}) looks like a smooth interpolation between the free-streaming
and hydrodynamic regimes, it does not mean that initially there are
no interactions and then interactions gradually switch on. One can see
that this is not the case from the fact that the expression
(\ref{green2}) is based on the approximate formal solution
(\ref{boltz-3-5}) of kinetic equation (\ref{boltz-1}).}
\begin{eqnarray}
\partial_{\mu}[(1-{\cal P}(x))T^{\mu
\nu}_{\text{hyd}}(x)]= - T^{\mu \nu}_{\text{free}}(x)
\partial_{\mu}{\cal P}(x). \label{tensor-eq}
\end{eqnarray}
The initial conditions for $T^{\mu \nu}_{\text{hyd}}$ follow from
the  right-hand side of  Eq.  (\ref{boltz-2}). That is, taking into
account approximation  (\ref{appr-2}) and utilizing the fact that initially
${\cal P}=1$, we see that the initial conditions are given by the
equation
\begin{eqnarray}
T^{\mu \nu}(x_{\text{in}})u_{\mu}(x_{\text{in}})=T^{\mu
\nu}_{\text{hyd}}(x_{\text{in}})u_{\mu}(x_{\text{in}}),
\label{initial-cond}
\end{eqnarray}
where $x_{\text{in}}\equiv(t_{0}(\textbf{r}),\textbf{r})$. Note that
the same equations defining the initial conditions for $T^{\mu
\nu}_{\text{hyd}}$ follow from  Eq. (\ref{boltz-1}). The solution of
Eq. (\ref{initial-cond}) is straightforward and follows from the
definition (\ref{011}):
\begin{equation}
\epsilon_{\text{hyd}}(x_{\text{in}})=\epsilon(x_{\text{in}}), \quad
u^{\mu}_{\text{hyd}}(x_{\text{in}})=u^{\mu}(x_{\text{in}})
\label{IC}
\end{equation}

As for the hydrodynamic pressure and equation of state (EOS),
$p_{\text{hyd}}= p_{\text{hyd}}(\epsilon_{\text{hyd}})$, they  can
be chosen, for example,  in agreement with the lattice QCD EOS for the RHIC
and LHC energies,  and the same for the viscous coefficients.  Then
one can use Eq. (\ref{tensor-eq}) with the initial conditions
(\ref{IC}) as phenomenological macroscopic equations for a
description of transient pre-hydrodynamic behavior in
ultrarelativistic heavy-ion collisions.

\section{A simple model to assess  the initial conditions for
hydrodynamics}

Let us begin with the example, which we choose merely because of its
simplicity, where the initial and final conditions are specified at
constant-time hypersurfaces $t=t_0$ and $t=t_{\text{th}}$,
respectively. Then, in obvious notations,
\begin{eqnarray}
p^{\mu}\frac{\partial}{\partial x^{\mu}} =
p_{0}\frac{\partial}{\partial t} +
\textbf{p}\frac{\partial}{\partial \textbf{r}},
 \label{ex-1}
\end{eqnarray}
and we get from (\ref{green1}) and (\ref{ex-1}) that
\begin{eqnarray}
G_{p}(x-x')=p_{0}^{-1}\Theta (t-t')\Theta (t'-t_0)
\delta^{(3)}(\mathbf{r}(t,t')-\textbf{r}'), \label{ex-2}
\end{eqnarray}
where $\Theta (t'-t_0)$ indicates that the evolution time starts at
$t_0$, and
\begin{eqnarray}
 \mathbf{r}(t,t')=\mathbf{r}-(\mathbf{p}/p_{0})(t-t')
\label{ex-2-1}
\end{eqnarray}
 satisfies  the equation
\begin{eqnarray}
p^{\mu}\partial_{\mu} \mathbf{r}(t,t')
 =0.
\label{ex-2-2}
\end{eqnarray}
 Substituting these results into (\ref{p-def}) and integrating over $\textbf{r}'$
 we get  an  expression that explicitly coincides with  (\ref{boltz-3-2}):
\begin{eqnarray}
{\cal P}(t,\textbf{r},p)=
\exp \left\{ -\stackrel{t}{%
%TCIMACRO{\underset{t^{\prime }}{\int }}%
%BeginExpansion
\mathrel{\mathop{\int }\limits_{t_0 }}}%
%EndExpansion
\frac{1}{t_{\text{rel}}(s,\mathbf{r}(t,s),p)}ds\right\},
\label{ex-2-3}
\end{eqnarray}
where
\begin{eqnarray}
t_{\text{rel}}(s,\mathbf{r}(t,s),p)= \frac{p_{0}\tau^*_{\text{rel}}
(s,\mathbf{r}(t,s),p)}{p^{\mu}u_{\mu}(s,\mathbf{r}(t,s))}.
 \label{ex-3}
\end{eqnarray}
Referring to Eqs.  (\ref{free-1}), (\ref{ex-1}) and (\ref{ex-2-2}),
we see that
\begin{eqnarray}
f_{\text{free}}(x,p)=f(t_0, \mathbf{r}(t,t_0),p). \label{ex-4-1}
\end{eqnarray}
Equation (\ref{ex-4-1}) allows to build the evolution of $T^{\mu
\nu}_{\text{free}}$ and after specification of $\tau^*_{\text{rel}}$
to fix the equation of relaxation (\ref{tensor-eq}). In the next
example, which is  very important in practice, we analyze this
procedure in detail.

Now let us consider the hypersurfaces $\tau=\tau_0$ and
$\tau=\tau_{\text{th}}$ as the initial and final ("thermalization")
hypersurfaces, respectively; here $\tau=\sqrt{t^{2}-z^{2}}$ is the
longitudinal proper time. In order to describe the boost-invariant
dynamics, it is convenient to switch to the coordinates
$(\tau,\textbf{r}_{T},\eta)$, where $\textbf{r}_{T}$ is thetransverse
radius vector and $\eta$ is the space-time rapidity, $\eta =\tanh
^{-1}z/t$; then $t=\tau \cosh \eta$ and $z=\tau \sinh \eta$. The
particle four-momentum can be expressed through the momentum
rapidity $y=\tanh ^{-1}p_{L}/p_{0}$ where $p_{L}$ is the longitudinal
momentum, the transverse momentum $\mathbf{p}_{T}$, and the transverse mass
$m_{T}=\sqrt{m^{2} + p_{T}^{2}}$, then $p^{\mu}=(m_{T}\cosh y,
\mathbf{p}_{T}, m_{T}\sinh y)$. Accounting for the boost-invariant
dynamics, we choose
\begin{eqnarray}
\tau_{rel}^*(x,p)=\tau_{rel}^*(\tau,\theta,{\bf r}_T,{\bf p}_T),
\label{ex-4-2}
\end{eqnarray}
where $\theta=y-\eta$. Transforming $p^{\mu}\partial_{\mu}$  to
these coordinates,  we find
\begin{eqnarray}
p^{\mu}\frac{\partial}{\partial x^{\mu}} = m_{T}\cosh \theta
\frac{\partial}{\partial \tau} + \frac{m_{T}}{\tau}\sinh \theta
\frac{\partial}{\partial \eta} +
\textbf{p}_{T}\frac{\partial}{\partial \textbf{r}_{T}}.
 \label{ex-5}
\end{eqnarray}
 The auxiliary function $G_{p}$ then reads
 \begin{eqnarray}
G_{p}(x-x')=\frac{\Theta (\tau - \tau') \Theta (\tau'-\tau_0)\delta
(\eta(\tau,\tau')-\eta')
\delta^{(2)}(\textbf{r}_{T}(\tau,\tau')-\textbf{r}')}{m_{T}
\cosh(y-\eta')}, \label{ex-6}
\end{eqnarray}
where $\Theta (\tau'-\tau_0)$ indicates that the time evolution
starts at nonzero proper time $\tau_0$. Here $\eta(\tau,\tau')$ and
$\textbf{r}_{T}(\tau,\tau')$   satisfy the equations
\begin{eqnarray}
p^{\mu}\partial_{\mu} \eta(\tau,\tau')=0, \label{ex-6-1} \\
p^{\mu}\partial_{\mu} \textbf{r}_{T}(\tau,\tau') =0, \label{ex-6-11}
\end{eqnarray}
and can be written as
\begin{eqnarray}
    \sinh \theta(\tau,\tau' )=\frac{\tau}{\tau'}\sinh\theta , \\  \label{ex-5-1}
    {\bf r}_T(\tau,\tau')={\bf r}_T-\frac{{\bf
    p}_T}{m_T}(\tau\cosh\theta-\sqrt{\tau'^2+\tau^2\sinh^2\theta}),
\label{ex-5-2}
\end{eqnarray}
where $\theta(\tau,\tau' ) = y- \eta(\tau,\tau')$. Substituting
(\ref{ex-6}) into (\ref{p-def}), accounting for (\ref{ex-4-2}), and
integrating over $\eta'$ and  $\textbf{r}'_{T}$, we get
\begin{eqnarray}
{\cal P}(\tau,\theta,{\bf r}_T,{\bf
p}_T)=\exp{\left(-\int\limits_{\tau_0}^{\tau
}\frac{1}{\tau_{\text{rel}}(\tau',\theta(\tau,\tau'),\textbf{r}_{T}
(\tau, \tau'),{\bf p}_{T})}d\tau'\right)},
 \label{ex-8}
\end{eqnarray}
where
\begin{eqnarray}
\tau_{\text{rel}}(\tau',\theta(\tau,\tau'),\textbf{r}_{T} (\tau,
\tau'),{\bf p}_T) =\frac{m_T \cosh\theta(\tau,\tau')}{p^\mu u_\mu
(\tau',\theta(\tau,\tau'),\textbf{r}_{T} (\tau,
\tau'))}\tau_{\text{rel}}^{*}
(\tau',\theta(\tau,\tau'),\textbf{r}_{T} (\tau, \tau'),{\bf p}_T).
 \label{ex-9}
\end{eqnarray}
Then $f_{\text{free}}(x,p)$ is given by
\begin{eqnarray}
f_{\text{free}}(x,p)=f(\tau_0,\theta(\tau,\tau_0),\textbf{r}_{T}
(\tau, \tau_0),{\bf p}_T). \label{ex-7}
\end{eqnarray}

In order to pass now to a description of the relaxation dynamics
at the prethermal stage, let us take into account that at this
stage the longitudinal flow is much stronger than the transverse
one and so, according to Eq. (\ref{ex-9}), $\tau_{\text{rel}}
\approx \tau_{\text{rel}}^{*} $. Therefore, to get a convenient
parametrization of the prethermal evolution of the
energy-momentum tensor of the system, one might assume that the
relaxation time $\tau_{\text{rel}}^{*}$ in the rest frames of the
fluid elements depends mainly only on  the proper time $\tau$:
$\tau_{\text{rel}}^{*} = \tau_{\text{rel}}^{*}(\tau)$. Then $
{\cal P}$ also depends  on $\tau$ only. It corresponds to the
Bjorken picture \cite{Bjorken}, where the thermalization processes
are supposed to be synchronous in proper time of the fluid
elements, so that the complete thermalization and the beginning of
the hydrodynamic expansion happen at some common proper time
$\tau_{\text{th}}$. Within such an approximation we get
\begin{eqnarray}
T^{\mu \nu}(x)=T^{\mu \nu}_{\text{free}}(x){\cal
P}(\tau)+T_{\text{hyd}}^{\mu \nu}(x)(1-{\cal P}(\tau)),
\label{ex-14}
\end{eqnarray}
 where
 \begin{eqnarray}
T_{\text{free}}^{\mu \nu}(x)=\int d^{2}p_T d\theta
p^{\mu}p^{\nu}f(\tau_0,\theta(\tau,\tau_0),\textbf{r}_{T} (\tau,
\tau_0),{\bf p}_T), \label{ex-13}
\end{eqnarray}
and $T_{\text{hyd}}^{\mu \nu}(x)$ is associated with ideal or
viscous hydrodynamics;  for example, for the former it can be parametrized as
follows:
\begin{eqnarray}
T^{\mu \nu}_{\text{hyd}}(x)= (\epsilon_{\text{hyd}}(x) +
p_{\text{hyd}}(x))u^{\mu}_{\text{hyd}}(x)u^{\nu}_{\text{hyd}}(x) -
p_{\text{hyd}}(x)g^{\mu \nu},
 \label{ex-17}
\end{eqnarray}
 where $\epsilon_{\text{hyd}}(x)$ is the energy density in the comoving system and $p_{\text{hyd}}(x)$ is
 the pressure. The corresponding evolutional equations are\footnote{Note  that Eq.
(\ref{ex-15}) can be considered in computational dynamics  as the
hydrodynamic equation for the energy-momentum tensor $\tilde{T}^{\mu
\nu}_{\text{hyd}}=(1-{\cal P}(\tau))T^{\mu \nu}_{\text{hyd}}$ of an
ideal fluid with an explicit "source" term on the right-hand side and
with rescaled energy density
$\tilde{\epsilon}_{\text{hyd}}=(1-{\cal
P}(\tau))\epsilon_{\text{hyd}}$, and pressure
$\tilde{p}_{\text{hyd}}=(1-{\cal P}(\tau))p_{\text{hyd}}$. Also,
since in our approximation the relaxation time and, hence, the
probability ${\cal P}$ of freely propagation are  supposed to be known
functions (up to the two fitting parameters), as well as the tensor
of free propagating system $T^{\mu \nu}_{\text{free}}(x)$, then Eqs. (\ref{ex-15}) are of  the same
 type as the equations
$\partial_{\mu}T^{\mu \nu}_{\text{hyd}}(x)=0$. So, if the latter are
equations of the hyperbolic type, as in the case of  the ideal fluid
or second-order viscous hydrodynamics (in the form of
Israel and Stewart), then  causality is preserved in our approach.}
\begin{eqnarray}
\partial_{\mu}[(1-{\cal P}(\tau))T^{\mu
\nu}_{\text{hyd}}(x)]= - T^{\mu \nu}_{\text{free}}(x)
\partial_{\mu}{\cal P}(\tau). \label{ex-15}
\end{eqnarray}
In order to link  Eqs. (\ref{ex-14}) and  (\ref{ex-15}) with the 
hydrodynamics at $\tau=\tau_{\text{th}}$, one needs 
\begin{eqnarray}
{\cal P}(\tau_0)=1, \quad  {\cal P}(\tau_{\text{th}})=0, \quad \partial_{\mu}{\cal
P}(\tau_{\text{th}})=0.
 \label{ex-20}
\end{eqnarray}
 The initial
conditions for $T^{\mu \nu}_{\text{free}}$ coincide with initial
conditions for $T^{\mu \nu}$, namely $T^{\mu
\nu}_{\text{free}}(\tau_{0},\textbf{r}_{T},\eta)=T^{\mu \nu}
(\tau_{0},\textbf{r}_{T},\eta)$. The initial conditions for $T^{\mu
\nu}_{\text{hyd}}$ (see Eq. (\ref{initial-cond})) are defined by Eq.
(\ref{IC}). The hydrodynamic pressure and equation of state,
$p_{\text{hyd}}= p_{\text{hyd}}(\epsilon_{\text{hyd}})$,  can be
chosen, for example,  in agreement with the lattice QCD EOS.

In order to apply the equation of relaxation dynamics (\ref{ex-15})
to calculate the space-time evolution of the energy-momentum tensor
toward the hydrodynamic one,  the function ${\cal P}(\tau)$
has to be specified. Without a knowledge of the specific
thermalization dynamics, we propose to make the approximation for this
as simple as possible with a minimal set of parameters. With that
end in view, we do not discuss here the most general case, but
propose the following simple ansatz  for ${\cal P}(\tau)$:
\begin{eqnarray}
{\cal P}(\tau)=
\exp \left\{ -\stackrel{\tau}{%
%TCIMACRO{\underset{t^{\prime }}{\int }}%
%BeginExpansion
\mathrel{\mathop{\int }\limits_{\tau_0}}%
%EndExpansion
}\frac{1}{\tau_{\text{rel}}(s)}ds\right\},
 \label{ex-21}
\end{eqnarray}
where
\begin{eqnarray}
\tau_{\text{rel}}(s)=\tau_{\text{rel}}(\tau_0)\frac{\tau_{\text{th}}-s}{\tau_{\text{th}}-\tau_0}.
 \label{ex-22}
\end{eqnarray}
Performing the integral in (\ref{ex-21}), we find that
\begin{eqnarray}
{\cal P}(\tau)=  \left (
\frac{\tau_{\text{th}}-\tau}{\tau_{\text{th}}-\tau_0}\right
)^{\frac{\tau_{\text{th}}-\tau_0}
 {\tau_{\text{rel}} (\tau_0)}}.
 \label{ex-23}
\end{eqnarray}
The conditions (\ref{ex-20}) require $\frac{\tau_{\text{th}}-\tau_0}
 {\tau_{\text{rel}}
 (\tau_0)}>1$ (then also ${\cal P}/\tau^*_{\text{rel}}\rightarrow 0$ at
$\tau \rightarrow \tau_{\text{th}}$) which is a constraint on the
model parameters: the lifetime of the
 prehydrodynamical period $(\tau_{\text{th}}-\tau_0)$ and the initial value of the relaxation-time
parameter $\tau_{\text{rel}} (\tau_0)$. The energy-momentum
relaxation equation  (\ref{ex-15}) together with the proposed
parametrization (\ref{ex-23}) for the probability function ${\cal
P}(\tau)$ can be used to assess
$T^{\mu\nu}(\tau_{\text{th}},\textbf{r}_{T},\eta)=T^{\mu\nu}_{\text{hyd}}(\tau_{\text{th}},\textbf{r}_{T},\eta)$
and so find the initial conditions (in particular, the energy density and
transverse flow) for hydrodynamic simulations of ultrarelativistic
heavy-ion collisions.

The heuristic method, which allows us to derive these macroscopic
equations,  is based on the Boltzmann equation in the relaxation-time
approximation, but, to our mind, its applicability area is not
restricted to a rarefied system  just as the form of the hydrodynamic
equations based on approximate solutions of the Boltzmann equations
near the local equilibrium distribution can be  applied for  a
description of dense fluids if they are near a local thermal
equilibrium. It is worth noting that  Eqs. (\ref{ex-15}),
(\ref{ex-21}), and (\ref{ex-22})  do not contain details of the
system's properties, except for the initial energy-momentum tensor and its
free evolution. The method
makes it possible to match the initial nonequilibrium state of
the system and the locally equilibrated one for various kinds
of system, such as partons, strings, and a glasma field.
 It
also allows one to match a far-from-equilibrium initial state
not only to the locally equilibrated one but also to the state associated with the initial conditions for viscous hydrodynamic
evolution of the QGP.   For the latter,  Eqs. (\ref{ex-15}) and (\ref{ex-23}) preserve their
form, but the energy-momentum tensor $T^{\mu \nu}_{\text{hyd}}(x)$
 differs from the simple form of  perfect hydrodynamics
(\ref{ex-17}) and corresponds to the tensor of  viscous
hydrodynamics. It should be noted that in this case Eq.
(\ref{ex-22}) loses its interpretation as the relaxation time to the
local equilibrium state (it tends to zero at $\tau \rightarrow
\tau_{\text{th}}$) and so the probability (\ref{ex-23}) plays  now
just the role of the interpolating function. Nevertheless, it allows
one to match smoothly the very initial state of the matter in
\textit{A }+ \textit{A} collisions with the initial conditions for
viscous hydrodynamics when (partial) thermalization is already
established; in other words,
to find the initial hydrodynamic parameters - energy density,
hydrodynamic velocity field, and so on - for the dissipative
evolution of the QGP in agreement with the model relaxation
dynamics and the conservation laws utilized at the prethermal
stage.

\section{Summary}
In this paper we presented a simple model of the early stage
in ultrarelativistic heavy-ion collisions. The model describes 
smooth conversion of the preequilibrium dynamics into hydrodynamics,
either perfect or viscous. Our phenomenological
approach is motivated by the Boltzmann equation in the
relaxation-time approximation, accounts for the energy and
momentum conservation laws, and contains two parameters:
the lifetime of the prehydrodynamic stage and the initial value
of the relaxation-time parameter. The preequilibrium evolution
is modeled by the continuous evolution of the energymomentum
tensor from the initial far-from-equilibrium state
to the perfect or viscous fluid form. Themodel allows the flows
and energy densities to be assessed at a supposed time of initialization
of hydrodynamics, which then can be used as the initial
condition for hydrodynamic simulations of the further evolution
of matter (QGP) in ultrarelativistic heavy-ion collisions.

\begin{acknowledgments}

The research was carried out within the scope of the EUREA: European
Ultra Relativistic Energies Agreement (European Research Group:
"Heavy ions at ultrarelativistic energies") and is supported by the
Fundamental Researches State Fund of Ukraine, Agreement No
F33/461-2009 with the Ministry for Education and Science of Ukraine.

\end{acknowledgments}

\end{document}